\documentclass[12pt,preprint]{aastex}

\shorttitle{Lensing in LCDM universe: Secondary matter}

\shortauthors{Wambsganss, Bode, \& Ostriker }

\begin{document}

\title{Gravitational lensing in a concordance LCDM universe: 
The importance of secondary matter along
	the line of sight 
	}
	%%%%in a concordance $\Lambda$CDM universe}

\author{Joachim Wambsganss$^1$, Paul Bode$^2$, and 
		Jeremiah P.~Ostriker$^{3,2}$}

\affil{$^1$ Institut f\"ur Physik, Universit\"at Potsdam, 
	14467 Potsdam, Germany}
\affil{$^2$ Dept. of Astrophysical Sciences, Princeton University,
    Princeton, NJ 08544}
\affil{$^3$ Institute of Astronomy, Cambridge University, 
	Madingley Road, Cambridge, UK}
\email{jkw@astro.physik.uni-potsdam.de,bode@astro.princeton.edu,
		jpo@ast.cam.ac.uk}

\begin{abstract}

To date, in almost all strong gravitational lensing analyses 
for modeling giant arc systems and multiple quasar images,
it has been assumed that all the deflecting matter is concentrated 
in one lens plane at a certain distance--- 
the thin lens approximation.  However,
in a few observed cases, lenses at more than
one redshift have been identified as contributing
to the image splitting.
% (EXAMPLES: Kundic, Koopmans ...).

Here we report on a quantitative investigation of the 
importance and frequency of significant multiple
lensing agents. 
We use multi-lens plane simulations to evaluate
how frequently two or more lens planes combined 
are essential for multiple imaging, as compared 
with the cases where a single lens plane alone 
provides enough focusing to be supercritical. 

We find that 
the fraction of cases for which more than one lens plane
contributes significantly to a multi-image lensing situation
is a strong function of source redshift. 
For sources at redshift unity, 95\% of lenses involve 
only a single mass concentration, but
for a more typical scenario with, e.g.,  a source at a redshift of 
$z_s = 3.8$, as many as 38\% of the strongly lensed quasars/arcs
occur because of a significant matter contribution from one or more
{\em additional} lens planes. 
In the  30\% to 40\% of cases when additional planes make a significant
contribution, the surface mass density of the
primary lens will be overestimated by about 15\% to 20\%,
if the additional contributions are not recognized.

\end{abstract}

\keywords{cosmology: gravitational lensing, arcs, quasars, galaxy clusters }

\section{Introduction}

By now of order 100 multiple quasar systems
are known (cf. Kochanek et al. 2004), as well as
more than
100 galaxy clusters which produce giant luminous arcs
(for a selection, see Table 2 of Wambsganss, Bode \& Ostriker, 2004).
%(REFERENCE).
Usually it is assumed that modeling the matter
distribution as a thin lens, i.e. putting all the
matter responsible for the light deflection into
one lens plane of zero thickness, 
is a good enough approximation
(Schneider, Ehlers \& Falco 1992). 
And indeed, 
most systems can be modeled well
using this simplification. However, 
for some multiply imaged quasars no good models can be found
within this framework (e.g. for the quadruple
system B1422+231, see Kormann et al. 1994),  
and
on the observational side there are instances where 
galaxies at two different redshifts have been
identified as contributing to the lensing
(e.g. B1422+231, see Tonry 1998, or
for B2114+022, see Augusto et al. 2001 and Chae, Mao \& Augusto 2001).

There is one {\it a priori}
reason to expect that the contribution of
multiple lens planes may be important. If we consider a representative
shell of the universe centered around redshift $z = 1$ with comoving
thickness $\Delta l = 160 h^{-1}$Mpc, and
examine the distribution of surface mass densities, we find that
the fraction of ray bundles encountering more 
than the critical surface mass density\footnote{
	$\Sigma_{\rm crit}=\frac{c^2}{4\pi G}
	\frac{D_{\rm s}}{D_{\rm d}D_{\rm ds}}$, where
	$G$ and $c$ are the gravitational constant and 
		the velocity of light,
	$D_{\rm s}$, $D_{\rm d}$, and $D_{\rm ds}$, 
	are the angular diameter distances observer-source,
	observer-lens, and lens-source,respectively.}
$\Sigma > \Sigma_{crit}$
(here evaluated for a source redshift of $z_s \approx 3.8$)
is 
only $f_{\Sigma > \Sigma_{crit}} \approx 7 \times 10^{-6}$
(and of course proportional to $\Delta l$). 
%
%
%
% 040422: these numbers evaluated from distribution of
%	pixel densities in files like: density_distribution_2_015.ps
%
%
%
%
But, more to the
point, this  fraction steeply declines with increasing $\Sigma$:
$\delta \ln f / \delta \ln \Sigma_{\Sigma} 
	\approx  -4.5$ at $\Sigma \approx \Sigma_{crit}$
(for more details see in Wambsganss, Bode \& Ostriker 2004b). 
Thus, there are many rays along which 
the surface mass density is just subcritical for each one that 
is supercritical. And for each slightly subcritical ray, a small
additional mass concentration along the line-of-sight can make a 
significant difference in its behavior.

So far no quantitative estimate exists on how
frequently a second matter clump along
the line of sight contributes significantly
to the light deflection caused by a 
near critical primary matter concentration, 
hence affecting
the image geometry as well as the intensity
ratios. 
%
% inserted 031218, after fax with jpo's comments
%
%The primary reason for expecting a significant contribution 
%from non-dominant lens planes is simply stated:
%The probability that a line of sight through a 
%single lens plane has s supercritical
%surface mass density  is very small ($\approx 10^{-4}$) and 
%the probability distribution of the surface mass density
%is very steeply declining for values near critical. 
%Thus for every case of a supercritical line of sight, 
%there are many cases where it is sub critical by 
%a small amount. 
In these instances, 
a relatively small ``boost"' from an overdense region   
elsewhere along the line of sight  will make this 
line-of-sight supercritical.
In fact, we find that for a source redshift of $z_S = 2.5$ approximately
30\% of the cases of strong lensing are produced by multiple disparate
matter planes, and averaged over all cases, the secondary 
matter contribution is 13.2\% $\pm$ 13.4\% of the critical density.

Here we present results based on multiple lens plane
rayshooting simulations, using very high resolution
LCDM N-body simulations for the 
pseudo-3D matter distribution.
By following light rays through many lens planes up to
high source redshift, we model the behavior
of realistic light bundles.
In Section 2 we 
briefly describe the underlying simulations and the method
we use to evaluate the importance of more than one lens plane.
In Section 3 we present our results on how important two, three
or more lens planes are for strong lensing of sources at different
redshifts. Summary and conclusion can be found in 
Section 4.

\section{Method}

We performed ray shooting simulations in order to quantitatively
determine strong and weak lensing properties  of a concordance
LCDM model, 
as described in detail  in Wambsganss, Bode \& Ostriker (2004b).
This cosmological model has
the following parameters:
matter content
$\Omega_{\mathrm M}=0.3$, 
cosmological constant 
$\Omega_\Lambda=0.7$, 
Hubble constant 
$H_0=70$ km/sec/Mpc, 
linear amplitude of mass fluctuations 
$\sigma_8$=0.95,
and 
primordial power spectral index 
$n_s$=1 (consistent with the 1$\sigma$ WMAP derived
cosmological parameters,  see Spergel et al. 2003, Table 2).
The simulation, carried out with the TPM code (Bode \& Ostriker 2003),
has a comoving side length of  
$L=320 h^{-1}$Mpc;
the cubic spline softening length was set to $\epsilon=3.2 h^{-1}$ kpc,
producing a ratio of    
box size to softening length of $L/\epsilon =10^5$.
We used $N=1024^3$ 
particles,
with the individual particle mass being
$m_{\rm p}=2.54\times 10^9 h^{-1}$ M$_\odot$,
%
%
% inserted 031218, after fax with jpo's comments
%
so that a large halo, similar to those which produce the
observed giant arcs, would be represented by of order $10^6$ particles
which is enough to allow a fair representation of 
both the inner cusp and of significant substructure.

We produced lens screens 
%The output was stored at 19 redshift values 
out to 
$z_L \approx 6.4$, the centers of the lens screens
correspond to comoving distances of 
$(80 + k \times 160) h^{-1}$Mpc, where $k=0,...,35$. 
The comoving average surface mass density of the lens planes
is $<\Sigma> 
	\approx 1.34 \times 10^7 h^2 M_\odot$kpc$^{-2} 
	\approx 0.00265 h^2$ g cm$^{-2}.$
%
%The physical surface mass density 
%(in grams/cm$^2$) decreases with lower redshift because
%we consider the expansion of the universe properly
%and hence cannot treat the lens planes in comoving units.
%The values of the average surface mass densities hence
%monotonically decrease from XXX
%grams/cm$^2$ at redshift XXX to 
%Each screen has a mass per unit area of XXX 
%XXX grams/cm$^2$ a the lowest redshift lens plane at  z = XXX.
%(!!!TO BE DONE!!!)
%
%
%
%
More details on the numerical scheme can be found in Wambsganss
et al. (2004b). A first result on the statistics of giant
luminous arcs was published as 
Wambsganss, Bode \& Ostriker (2004a). 

The following analysis was carried out
in order to evaluate the importance of secondary, tertiary etc.
lens planes.
In each ray shooting run, we use a grid of 800$^2$ rays
to cover an area of about 20 arcmin on a side.  Each
ray with starting position $(i,j)$ is followed backward to
a given source redshift (we used 7 different values: $z_s =$ 0.5, 1.0,
1.5, 2.5, 3.7, 4.8, and 7.5).  At each of the lens screens 
$k = 0,\dots,35$ we determined the surface mass density
of the matter pixels 
in units of the critical surface mass density at that redshift:
$\kappa (i,j,k) = \Sigma(i,j,k) / \Sigma_{crit}(z_s,k).$
%
%
%$\Sigma_{\rm crit}=\frac{c^2}{4\pi G}
	%\frac{D_{\rm s}}{D_{\rm d}D_{\rm ds}}$
%($G,c$ are the gravitational constant and the velocity of light,
	%$D_{\rm s}$, $D_{\rm d}$, and $D_{\rm ds}$, respectively,
	%are the angular diameter distances observer-source,
	%observer-lens, and lens-source).
Then we identified the highest value of $\kappa (i,j,k)$ for 
fixed angular position $(i,j)$. 
If this value was above unity (``supercritical''),
that particular lens plane alone would be enough to produce strong
lensing/multiple imaging at this particular position.
If this condition was not fulfilled 
($\kappa (i,j,k) < 1$ for all $k$ out to the source), 
we checked whether the sum of the two 
highest values of $\kappa$ along this line combined
would exceed the critical value.  If so,
this would imply that the combination of
two planes could produce multiple imaging, whereas each individual
plane still is sub-critical. 
If this was not the case either, we
did the same exercise for the combination of the three, 
four, and five highest surface mass density values along 
this particular ray position $(i,j)$.
We did this for all $800^2$ angular positions in 
100 different realizations, and for all the seven
source redshifts.

\section{Results}
%\section{Results and Discussion: How important are secondary  lens planes?}

\begin{figure}[htb]
\plotone{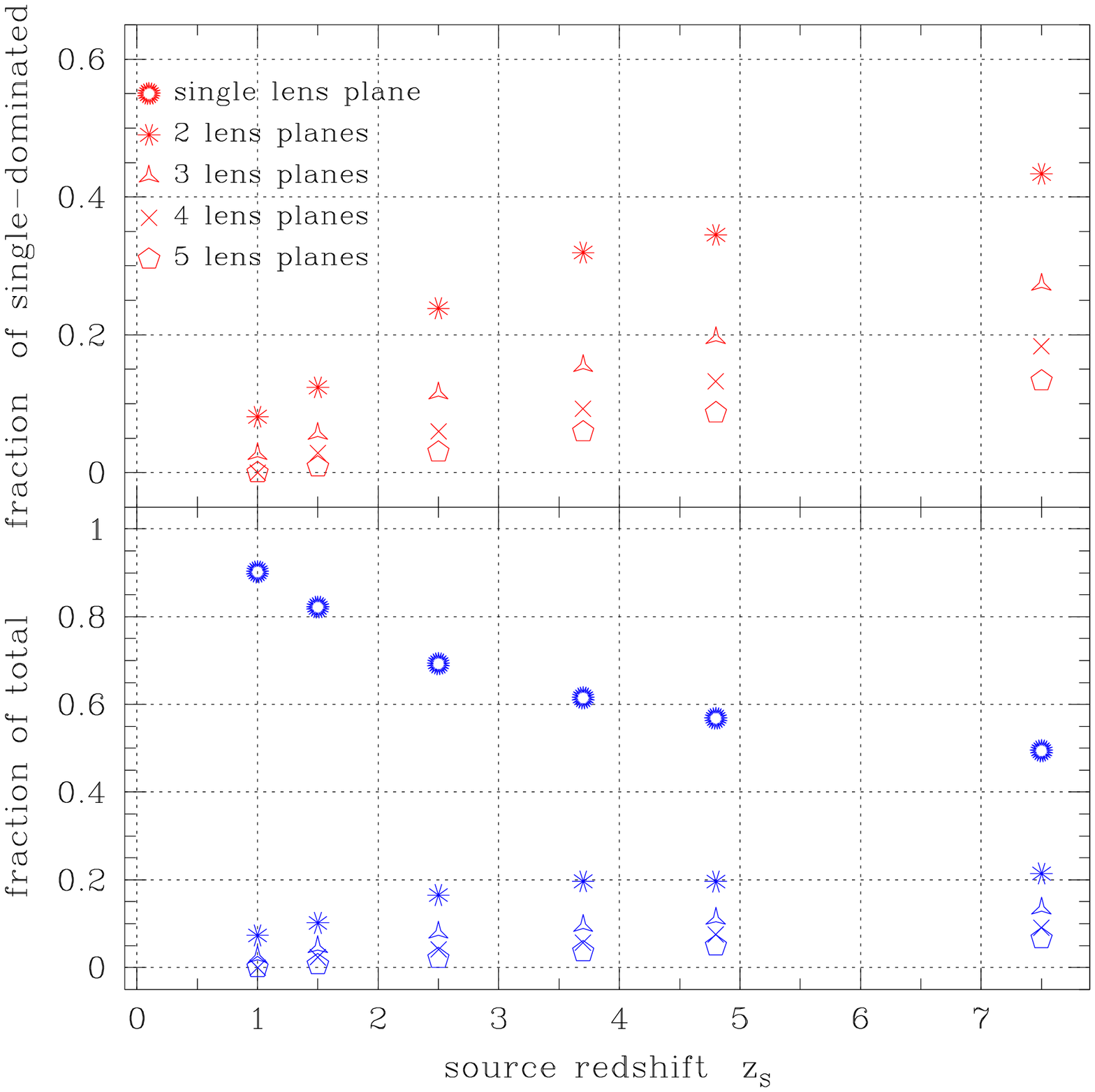}
\caption{Fraction of lines-of-sight for which a second (star),
a third (triangle), a fourth (square) and a fifth (pentagon)
lens plane contributes significantly
to the strong lensing in the sense that only
the value of the
combined surface mass density (sum of highest
two, three, four and five lens planes, respectively) 
is above critical and can produce
strong lensing/multiple images. 
The quantitative importance is
expressed as the ratio of the frequency of such cases
to the frequency of single plane lensing cases (top panel),
or of all strong lens cases (bottom panel).
%The quantitative importance of these cases is
%expressed as ``fraction'' of such cases compared
%to the cases of single-lens plane lensing (top panel),
%or compared to all strong lens cases (bottom panel).
These fractions are shown for the 
seven values of the source redshifts that we had considered:
$z_s =$ 0.5, 1.0, 1.5, 2.5, 3.7, 4.8, 7.5.
The thick circles in the bottom panel indicate the occurrence 
of single-lens plane lensing as a fraction of the total
strong lensing cases.
}
\label{fig_crit_kappa}
\end{figure}

In Figure  \ref{fig_crit_kappa} the fraction of lines-of-sight 
for which a second (third, fourth, fifth) lens plane 
contributes to reaching
the critical value of the surface mass density is 
displayed. 
The top panel indicates the contribution of secondary, tertiary etc. 
planes, expressed as a fraction of the frequency for which a single plane is 
supercritical; the bottom panel shows this 
contribution as the fraction  of {\em all} multiple
image cases,  including the `single plane' cases
(thus the symbols in the bottom panel add upp to unity).
 
The star symbol in Figure  \ref{fig_crit_kappa} 
indicates the situation in which two planes combined exceed
the critical surface mass density. This fraction is slowly increasing for
increasing source redshift: 
for $z_s = 1.0$, lines-of-sight in which two planes combined are super-critical 
are about 8\% compared to the single-lens plane cases. This fraction, however, 
monotonically increases to over 20\% for $z_s \ge 2.5$ and reaches 
more than 40\% for $z_s = 7.5$.
The other symbols indicate cases in which three (triangle), four (cross) and
five (pentagon) lines-of-sight combined exceed 
the critical value of the
surface mass density. 
%%%%The total number of cases declines with ``higher order": 
The overall importance of additional planes goes down, the
larger the number of significantly contributing lens planes is;
but in all cases the contribution is monotonically
increasing with source redshift. 
In the bottom panel of Figure  \ref{fig_crit_kappa}, the thick circles
indicate the fraction of all supercritical lines-of-sight which are
produced by a single lens plane: it starts as high as  95\% for
a source redshift of $z_s = 1.0$, and then it drops monotonically to
a value of 50\% for the highest source redshift of $z_s = 7.5$.

\begin{figure}[htb]
\plotone{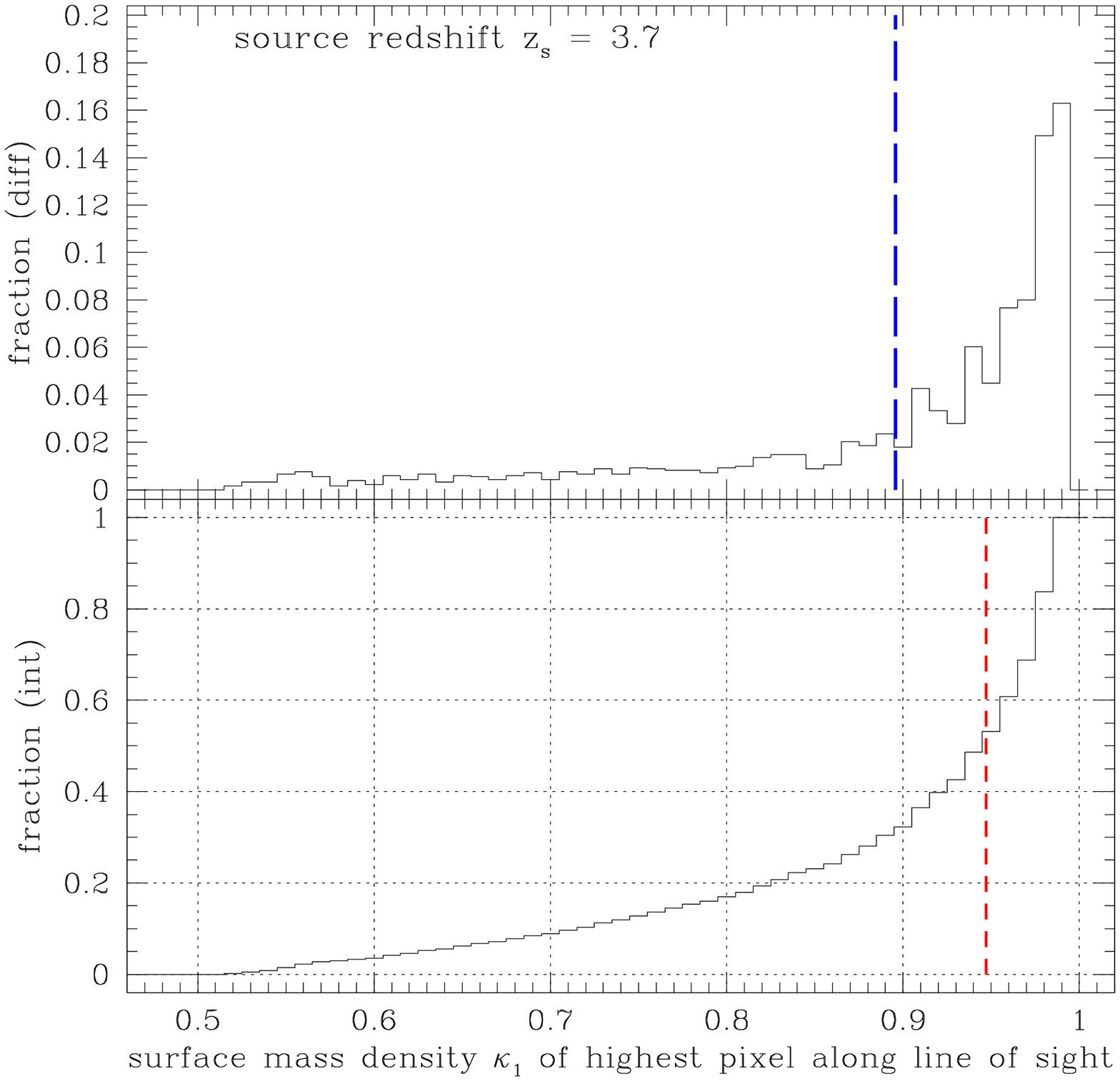}
\caption{Histogram of the distribution of 
the highest surface mass density value $\kappa_1$ along
the line of sight for cases in which 
only the combination of {\em two lens planes} 
produces super-critical surface mass density values
($\kappa_1 + \kappa_2 \ge \kappa_{\mathrm{crit}}$), 
for a source redshift of $z_s = 3.7$.
Top: differential distribution, the long-dashed vertical
line indicates
the {\em average} value $<\kappa_1> = 0.897$.
Bottom: integrated distribution, the short-dashed vertical
line shows 
the median value of the $\kappa_1$-distribution: 0.947.
}
\label{fig_over_2}
\end{figure}

In cases where two lens planes combined
to reach the critical surface mass density level, what is the relative
contribution of the two?
In Figure  \ref{fig_over_2}  this question is answered
for a source redshift of $z = 3.7$.
The histogram in the top panel shows the relative distribution of the
{\em primary} lens plane.  It is very low and flat for values of $\kappa \le 0.9$
(by definition, the ``primary'' plane in this case 
has to have $\kappa_1 \ge 0.5$).
Then the histogram steepens with the peak at $\kappa_1$-values 
just below
unity. 
This means even in a situation in which two lens planes are necessary
for critical lensing, 
the large majority of cases are dominated by one lens plane. 
The mean of the highest surface mass density
value for cases in which two planes contribute is: 
$<\kappa_1> \approx 0.90$.
The bottom panel shows the integrated fraction of these
cases in which the primary plane have 
$\kappa_1 \le  \kappa_{\mathrm{crit}}$,
but
$\kappa_1 + \kappa_2  \ge  \kappa_{\mathrm{crit}}$. 
It is easy to read off the median: 
in 50\% of the cases in which two planes contribute, the primary plane
has $\kappa_1 \ge  0.95$.

\begin{figure}[htb]
\plotone{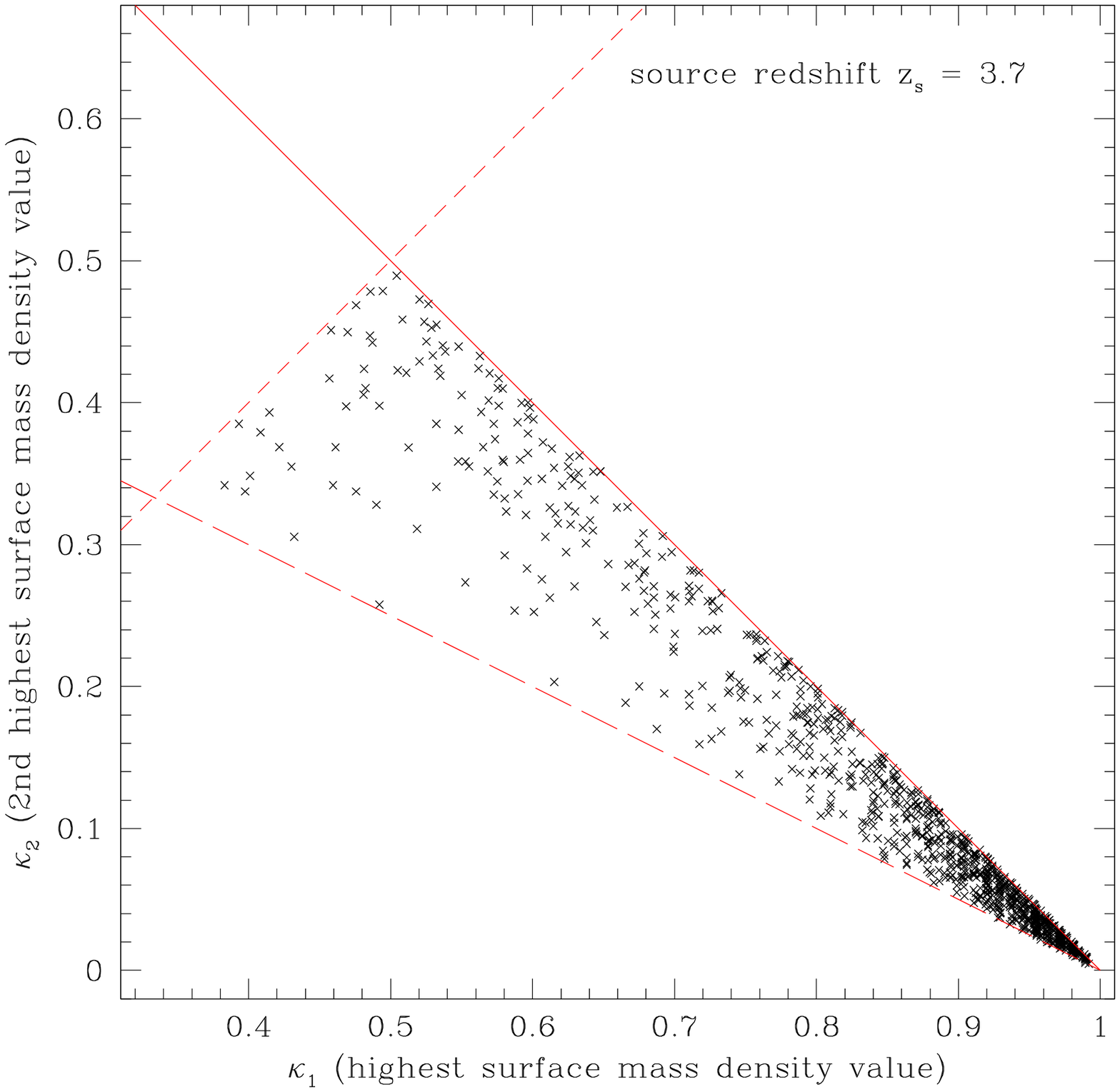}
\caption{
Each point in this diagram (for source redshift $z_s = 3.7$)
reflects one case for which the combination of 
{\em three lens 
planes} made this particular line of sight super-critical.
The location of the point indicates the values $\kappa_1$ and
$\kappa_2$, the highest and second highest surface mass density
value along the line of sight, respectively. 
The vertical distance of each point to the
solid line $\kappa_1 + \kappa_2 = 1$ indicates the minimum value
of the third contribution, $\kappa_3 \ge 1 - (\kappa_1 + \kappa_2)$. 
The dashed lines mark the additional
boundaries for the location of the points:
All points have to lie below the short-dashed 
line  $\kappa_2 = \kappa_1$ (because $\kappa_1 \ge \kappa_2$ 
by definition), and 
above the long dashed line
$\kappa_1 = 1 -  2 \times \kappa_2$ (because
$\kappa_3 \le \kappa_2$ by definition,
and $\kappa_2 + \kappa_3 \ge 1 - \kappa_1$ by requirement).
}
\label{fig_over_3}
\end{figure}

The same question can be asked about
cases in which {\em three} planes combined reached the
critical value of the surface mass density. In 
Figure  \ref{fig_over_3},  
each point represents one pair ($\kappa_1$,$\kappa_2$), i.e. the highest
and the second highest value of the surface mass density along the line
of sight. 
%The line marks the value 		%$\kappa_{\mathrm tot} =  1.0$, 
All points fulfill the criterion 
$\kappa_1 + \kappa_2 <  1.0$, and hence 
have to be left of the solid line. One can 
read off the (minimum) value of the third contribution $\kappa_3$
by the vertical distance of each point to this line. 
Most points are concentrated close to the solid line, which
means in most cases the third plane contribution is small.
The distribution of points
is limited towards  the top left by the relation
$\kappa_1 \ge  \kappa_2$ (short-dashed line).
Points close to this line represent cases for which
the two highest matter concentrations along the
line of sight are comparable.
The long-dashed line limiting the distribution
towards the bottom is given by the 
relation $\kappa_1 + 2 \times \kappa_2$ = 1: the secondary
and tertiary contributions are nearly equal for points near
this line.
The overall distribution of the
points clearly have the highest density towards the 
lower right, which
means that in most of the ``three-planes-contribute''  cases, one plane
clearly dominates as well.

\begin{figure}[htb]
\plotone{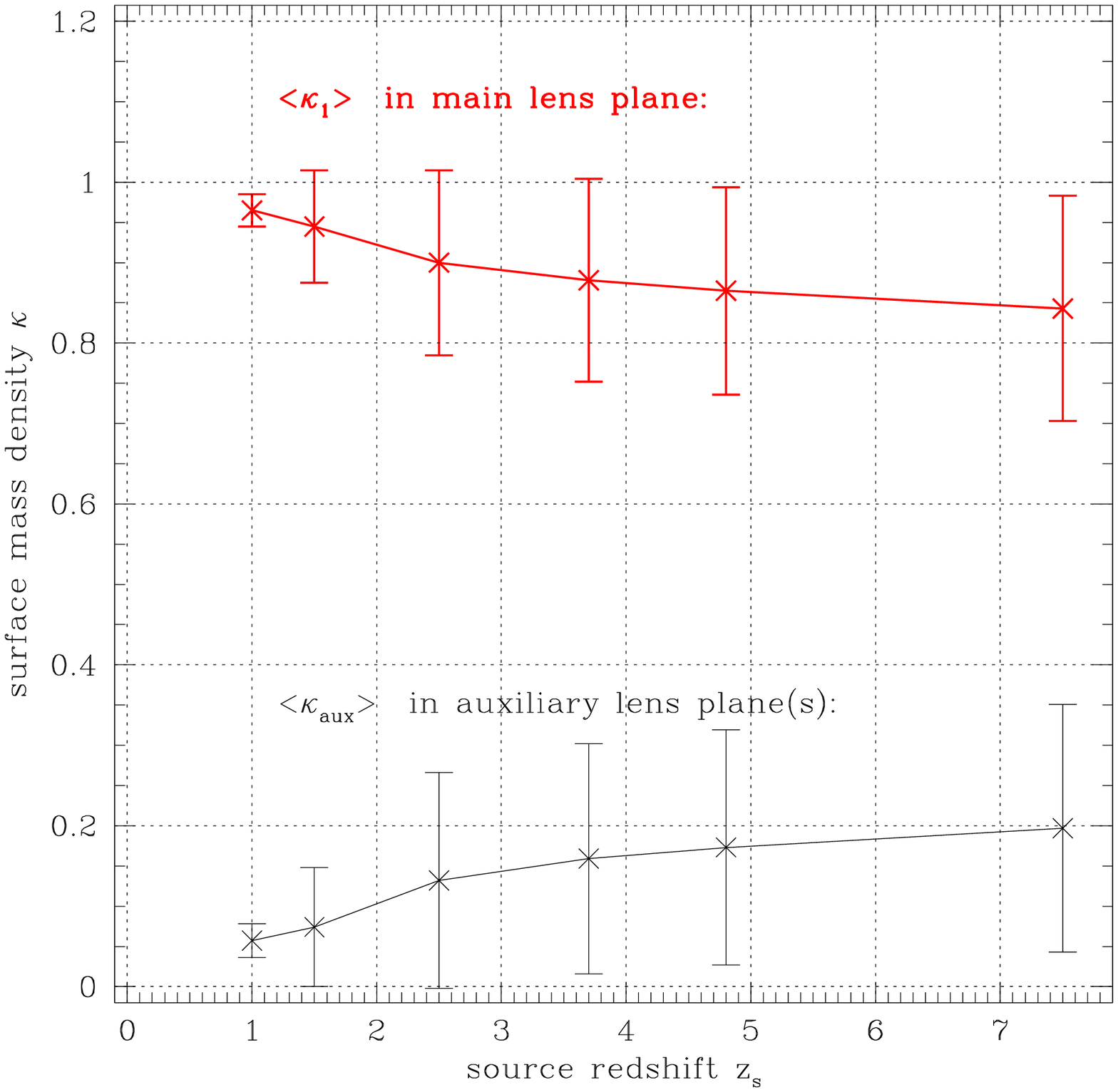}
\caption{Average surface mass density 
$<\kappa_1>$
in the dominant lens plane
(top points/line), 
and in the 
significantly contributing additional lens planes
($<\kappa_{\mathrm{aux}}>$, bottom points/line)
for cases in which two or more lens planes are
required to reach the critical surface mass density 
for strong lensing 
(error bars indicate the rms-fluctuations).
}
\label{fig_over_n}
\end{figure}

%%Another way to look at the contribution and importance
%%of auxiliary lens planes is the following.
Given an observed strong lensing situation, 
the probability that a single lens plane alone has a 
super-critical surface mass density
can be read off from the bottom part of Figure 
\ref{fig_crit_kappa} as a function of source redshift. 
An interesting question concerns a variant of this:
what is the average contribution of the secondary
lens planes to the surface mass density
$\kappa$ in those cases in which two, three or more lens planes
contribute significantly? 
This question can be answered by looking at
Figure \ref{fig_over_n}. 
Here only the cases in which two or more lens planes
combined reach the critical surface mass density are shown.
The two sets of lines show
the average surface mass density
of the dominant lens plane including its dispersion as
a function of source redshift (top, between
$0.8 \le \ \ <\kappa_1> \ \ \le 1.0$), and 
the combined average surface
mass density of the secondary, tertiary or higher order lens 
planes which
combined with the dominant one make this line of sight
supercritical (bottom, between 
$0.05 \le \ \ <\kappa_{\mathrm{aux}}> \ \ \le 0.2$).
A single mass concentration is clearly dominating.
Thus, in the typical case, if a cluster of galaxies is known to be
acting as a lens, its surface mass density will be overestimated
by 15\% to 20\% in the 30\% to 40\% of the cases when two or
more lens planes contribute significantly.

\section{Discussion }
%{\tt 
	%(the following typed in by jkw on March 3, after handwritten
	%comments from jpo on Jan. 24 or so)
%}
A small fraction of the  sky  is covered by regions that have a
surface mass density equal to the cosmic value 
$\Sigma_{\mathrm{crit}}$
		%
		%hxh \approx ???  (H_0 c /G)$ 
		%
capable of producing multiple images,
and this fraction declines rapidly with increasing 
surface mass density $\Sigma$, for $\Sigma \approx \Sigma_{crit}$. 
Thus many regions that
are subcritical will be boosted over the threshold by additional
chance accumulations of mass at unrelated distances,
which intersect the line of sight defined by: 
observer $\rightarrow$ main lens $\rightarrow$ source.
For source redshifts above $z = 2.5$  this occurs in
30\% to 40\% of the cases, and in these cases on average 
15\% to  20\% of the lensing mass is in the chance alignments.

This phenomenon has severe consequences which  must be considered
when using strong lenses as a tool in cosmological investigations.
Primary among them are the following:
\begin{enumerate}
\item the probability of strong lensing increasing with source 
	redshift more steeply than would be computed by the single
	sheet approximation;
\item masses of individual clusters will be overestimated by
	gravitational lensing techniques as compared to 
	%(for example)
	those using internal cluster velocity dispersions and/or 
	X-ray temperatures;
\item substructure may be incorrectly inferred to exist within
	lensing systems and time delays incorrectly computed.
\end{enumerate}
While the effects are not large, they are systematic and lensing
systems should be analyzed with consideration of these 
gravitational contaminations.

%\section{Summary}
%INS3 
As a consequence, 
these effects will lead to a  systematic error in the
analyses of lensing clusters such that the ``best fit'' single
lens plane gravitational lens models may overestimate the cluster mass 
by up to 10\% (depending in detail on lens and source
redshifts). 
Correspondingly this effect may introduce a systematic 
bias in the
determination of 
the Hubble constant 
of quasar lens systems which is usually 
based on the ``thin lens approximation''
(cf. Kochanek 2003, and Kochanek \& Schechter 2004).
%
%on an ensemble of 
%it may lead to a 
%typical underestimate
%%in the derived Hubble constant (RED) by the same order of
%magnitude. 
%{\tt original sentence from jpo: 
%We will not further
%analyze the implications unrecognized secondary mass contributions in 
%this paper.}
%
Further and more quantitative investigations of the implications of 
unrecognized secondary mass contributions 
are beyond the scope of this letter and 
will be done in subsequent analyses.

\acknowledgments

This research was supported by the National Computational Science
Alliance under NSF Cooperative Agreement ASC97-40300, PACI Subaward 766;
also by NASA/GSFC (NAG5-9284).  Computer time was provided by NCSA
and the Pittsburgh Supercomputing Center.

\end{document}